# GAS DYNAMIC THEORY OF LOCAL QUASIGRAVITY

## V.T. Volov

Samara Scientific Center of Russian Academy of Sciences,
3a, Studenchesky Lane, Samara 443001
Tel./fax. (846) 995-21-61, e-mail: vtvolov@mail.ru

## Abstract

In the present work there was found a class of noninertial frames of reference, which satisfy Einstein "equivalency" principle more than the known noninertial frames – these are strongly swirling gaseous flows.

Field intensity and potential in the mentioned frames of reference are similar to the corresponding values of natural gravity fields, but have the opposite sign. Scalar curvature of this space is negative and proportional to absolute gas temperature $T^*$.

There was obtained a new solution of Einstein equation which refers to type I in Petrov's classification for cylindrical symmetrical swirling ideal gas with variable angular velocity and nonzero pressure. The equation of state has a more complicated form $p_{quasi} = f\left(\varepsilon, T^*_{gas}, P^*_{gas}, k\right)$ than the known equations of state in theory of the vacuum.

1. **Introduction**

A. Einstein "equivalency" principle declares the similarity between gravity fields and noninertial frames of reference. Bodies will move in a similar way, i.e. they will have the same acceleration as while moving in a gravity field, caused by the distribution of a certain mass in a space [1]. However, the real equity of natural gravity and "effective" gravity fields, generated by noninertial frames of reference, can only be true locally for stationary fields [1, 2]. This is explained first of all by the degeneration of a natural gravity field in infinity, at the same time, the fields generated by noninertial systems, on the contrary grow unrestrictedly in infinity (rotary frames of reference), or stay finite in value for noninertial frames of reference that move forward with constant acceleration.

Thus, fields produced by noninertial frames of reference, are equivalent to genuine gravity fields, generated by a mass, distributed in space in a narrow space-time scale, where gravity field can be considered homogeneous. Every noninertial system can be opposed to another noninertial one, with regard to which all gravity effects disappear. The above listed differences between genuine gravity fields and noninertial fields are reflected in the analytical description of space-time properties [1]. However, there is a class of noninertial systems that meets equivalency principle more than other noninertial frames – it is swirling gaseous flows.

The structure of the present article consists of Introduction (I), Section II where Einstein equivalency principle and a class of noninertial frames of reference are analyzed, in Section III the solution of Einstein equations is given for cylindrical symmetrical swirling ideal gas flow with variable angular velocity, Section IV presents particular features of quasigravity fields analysis while using mathematical apparatus of exterior differential forms [11, 12], Section V contains conclusions and prospects of further research and a list of references.

## II. Einstein equivalency principle and one class of noninertial frames of reference

Equations of a viscous gas flow in cylindrical coordinate system $(r, \varphi, z)$ for stationary axially symmetric case ($\partial/\partial t = \partial/\partial \varphi \equiv 0$) look as follows:

$$V_r \frac{\partial V_r}{\partial r} + V_z \frac{\partial V_r}{\partial z} - \frac{V_\theta^2}{r} = -\frac{1}{\rho}\frac{\partial P}{\partial r} + \nu\left(\frac{\partial^2 V_r}{\partial r^2} + \frac{\partial^2 V_r}{\partial z^2} + \frac{1}{r}\frac{\partial V_r}{\partial r} - \frac{V_r}{r^2}\right); \qquad (1)$$

$$V_r \frac{\partial V_\varphi}{\partial r} + V_z \frac{\partial V_\varphi}{\partial z} + V_r \frac{V_\varphi}{r} = \nu\left(\frac{\partial^2 V_\varphi}{\partial r^2} + \frac{\partial^2 V_\varphi}{\partial z^2} + \frac{1}{r}\frac{\partial V_\varphi}{\partial r} - \frac{V_\varphi}{r^2}\right); \qquad (2)$$

$$V_r \frac{\partial V_z}{\partial r} + V_z \frac{\partial V_z}{\partial r} = -\frac{1}{\rho}\frac{\partial P}{\partial z} + \nu\left(\frac{\partial^2 V_z}{\partial r^2} + \frac{\partial^2 V_z}{\partial z^2} + \frac{1}{r}\frac{\partial V_z}{\partial r}\right). \qquad (3)$$

Equations of through flow, flow energy and state will accordingly look like:

$$\frac{\partial(\rho \cdot rV_r)}{\partial r} + \frac{\partial(\rho \cdot rV_z)}{\partial z} = 0; \qquad (4)$$

$$\rho\left(V_r \frac{\partial T}{\partial r} + V_z \frac{\partial T}{\partial z}\right) = V_r \frac{\partial p}{\partial r} + V_z \frac{\partial p}{\partial z} + \frac{1}{r}\frac{\partial}{\partial r}\left(r\lambda \frac{\partial T}{\partial r}\right) + \frac{\partial}{\partial z}\left(\lambda \frac{\partial T}{\partial z}\right) + Diss\Phi; \qquad (5)$$

$$p = \rho \frac{R}{\mu} T; \qquad (6)$$

$Diss\Phi =$

$$= \mu'\left\{2\left[\left(\frac{\partial V_r}{\partial r}\right)^2 + \left(\frac{V_r}{r}\right)^2 + \left(\frac{\partial V_z}{\partial z}\right)^2\right] + \left(\frac{\partial V_r}{\partial z} + \frac{\partial V_z}{\partial r}\right)^2 + \left(\frac{\partial V_\varphi}{\partial r} - \frac{V_\varphi}{r}\right)^2 - \frac{2\mu'}{3}\left(\frac{\partial V_r}{\partial r} + \frac{V_r}{r} - \frac{\partial V_z}{\partial z}\right)^2\right\},$$

where $\rho$ is a gas density, $V_r, V_\varphi, V_z$ - velocity constituents, $p$ - pressure, $\nu, \lambda$ - kinematic viscosity and thermal conduction of the gas, $R_G$ - universal gas constant, $\mu'$ - dynamic gas viscosity, $Diss\Phi$ - dissipative function.

Equations (1 ÷ 4) are complemented with boundary conditions and the dependence of thermophysical quantities on thermodynamic parameters. As it follows from the motion equation analysis (2), the solution of motion equation for whirl velocity constituent should be the law of a solid body rotation $(V_\varphi = \omega \cdot r)$ and potential gas flow $(V_\varphi \sim 1/r)$. The given solutions are common for ideal gas and special for viscous laminar gas flow.

Another specific solution of Navier-Stokes equation [3], regular in terminal cylinder stationary solution, often called Rankine or Burgers vortex, also has distinct areas of a flow core rotation ($\omega \approx const$) and potential flow $(V_\varphi \sim 1/r)$ (Fig.1). This fact proves the stability of the given solutions. The experiment confirms this statement – for swirling flows in chambers (Fig.2, 3) and for free jets (Fig.4) [4] there are two distinctly marked flow areas: 1) solid body rotation area and 2) potential flow. It should be noted that finiteness of the whirl velocity field generated by swirling gas-liquid flows in chambers occurs on a chamber wall due to gas adherence.

The research papers [4 ÷ 8] show that in the analysis of strongly swirling flows in swirl chambers, where the record values of turbulent kinematic viscosity take place $(v_{turb} = v_{mol} \cdot 10^4)$, the neglect of viscosity effects is justified by the circumstance that inertial powers play more important role than the power of viscosity, except for narrow wall boundary layer. That is why the use of ideal gas equations for the analysis of thermal gas dynamic parameters of strongly swirling gas flow gives enough accuracy. Thus, neither gas viscosity nor its compressibility or chamber walls can change significantly the character of strongly swirling gas flows.

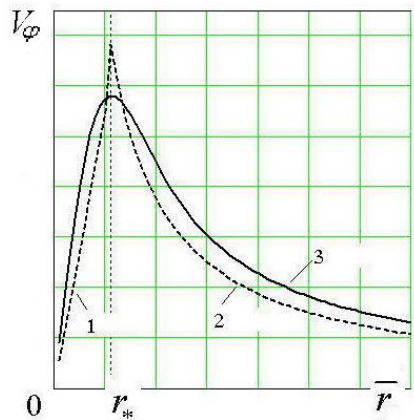

Fig.1.
Changing of whirl velocity constituent $V_\varphi$ from radius $r$

1- the law of a solid body rotation $V_\varphi = \omega \cdot r$; 2- the law of potential flow $V_\varphi = \dfrac{V_{\varphi 0} r_0}{r}$ ; 3- Burgers vortex $V_\varphi = \dfrac{\gamma}{2\pi \cdot r}\left(1 - \exp(-v \cdot r^2)\right)$ ; $v$ -gas viscosity; $\gamma = 2\pi r_0 V_{\varphi 0}$ - gas circulation.

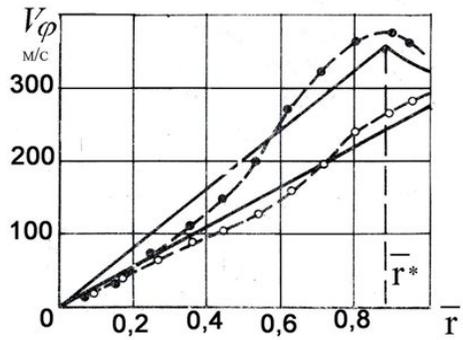

Fig.2.
Dependence of swirling flow tangent velocity on the radius of a vortex chamber

$\pi^* = P_1^*/P_{ax}$ - total degree of gas expansion in a vortex; $P_1^*$ - total pressure at a vortex chamber inlet; $P_{ax}$ - static pressure on a vortex chamber axis; ○ ● - the data [6]; ▬ calculated value of tangent velocity in a vortex chamber for ideal gas ($V_\varphi = \omega \cdot r$, $r < r_*$ и $V_\varphi = const/r$, $r > r_*$).

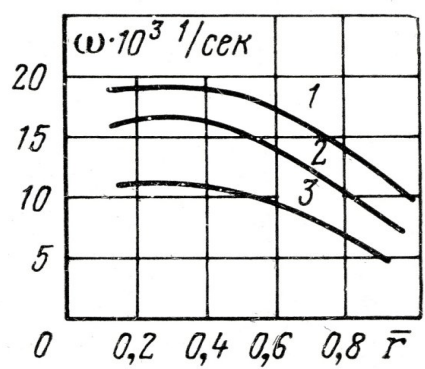

ig.3.
Dependence of a gas flow angular velocity in a vortex chamber on a chamber relative radius according to the data [6].

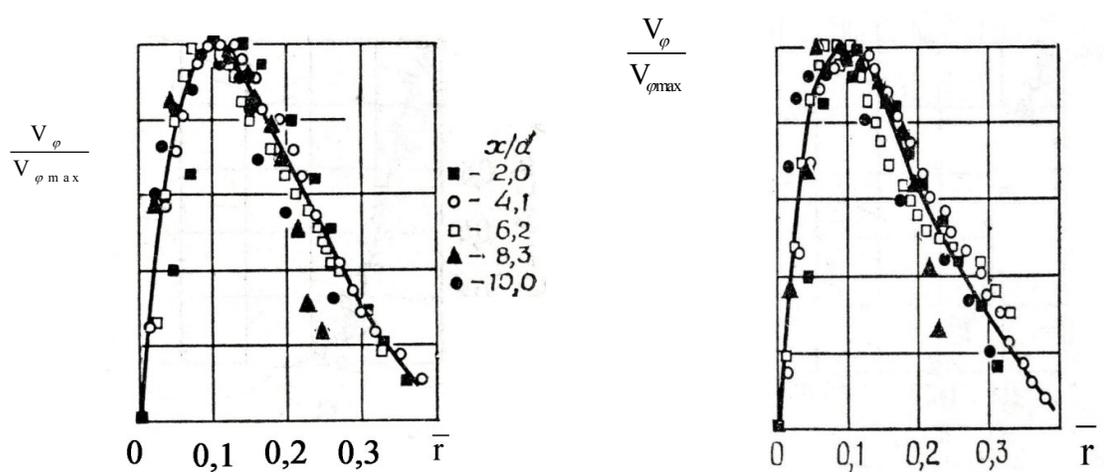

Fig.4.

Dependence of a gas relative tangent velocity in freely swirling jets on a chamber relative radius $r$

■, ○, □, ▲, ● - experimental findings [4].

Using the same approach let us consider modeling of quasigravity fields by ideal swirling gas flows for cylindrical symmetrical stationary gas flow ($\partial_t = \partial_\varphi = \partial_z = 0$). Here the term "quasigravity fields" is introduced to emphasize the difference of these fields, brought about by strongly swirling gas flows, from real gravity fields caused by mass distribution in the space.

By "quasigravity field" we mean a field that has potential and intensity distribution qualitatively similar to that of a real gravity field, but decreasing quicker than ($1/r$) and ($1/r^2$) correspondingly for potential $\varphi$ and field intensity $g$ in exterior zone ($r > r_*$).

In the core of a strongly swirling ideal compressible gas flow, as noted above, the law of a solid body rotation works, that is why quasigravity field intensity and potential change according to the following laws:

$$\begin{cases} \vec{g}_{quasi} = \omega^2 \vec{r}; \\ \varphi_{quasi} = \omega^2 \left( r_*^2 - \dfrac{r^2}{2} \right); \end{cases} \quad \begin{array}{l} \left| \vec{g}_{quasi} \right| << \vec{g}_{natural}; \\ \left| \varphi_{quasi} \right| < \varphi_{natural} \approx Const. \end{array} \quad r \leq r_*, \qquad (7)$$

where $\omega$ - angular velocity of a gas flow core rotation, $r_*$ - radius of flow separation into a swirl zone ($\omega = const$) and potential flow area ($V_\varphi \sim 1/r$), $g_{body,gas}, \varphi_{body,gas}$ - intensity and potential of a natural field, generated by graviting body and mass of gas.

The potential of a natural gravity field $\varphi_{body,gas}$ within the research given is practically constant and is not therefore considered when calculating quasigravity field.

For ideal compressible gas we can analytically find the radius of separation $r_*$ from the condition of pressure equity in the zone of flow sewing [6-7]; we can also find the flow velocity at a swirl chamber inlet $V_\varphi$ and the maximum tangential velocity value $V_{\varphi \max}$, that lets us work out the value of the flow core angular velocity ($\omega = V_{\varphi \max}/r_*$) depending on mode, geometrical parameters of swirling flow and gas physic-chemical properties.

Quasigravity field intensity and potential outside the flow core can be described with the following expressions:

$$\begin{cases} \vec{g}_{quasi} = \dfrac{\omega^2 r_*^4}{r^3} \\ \vec{\varphi}_{quasi} = \dfrac{\omega^2 r_*^4}{2r^2} \end{cases} \quad r > r_*. \qquad (8)$$

The analysis of natural gravity fields generated by mass distribution in a space and of quasigravity fields, caused by swirling gas flows makes it possible to draw a conclusion about their similarity (Fig. 5, 6).

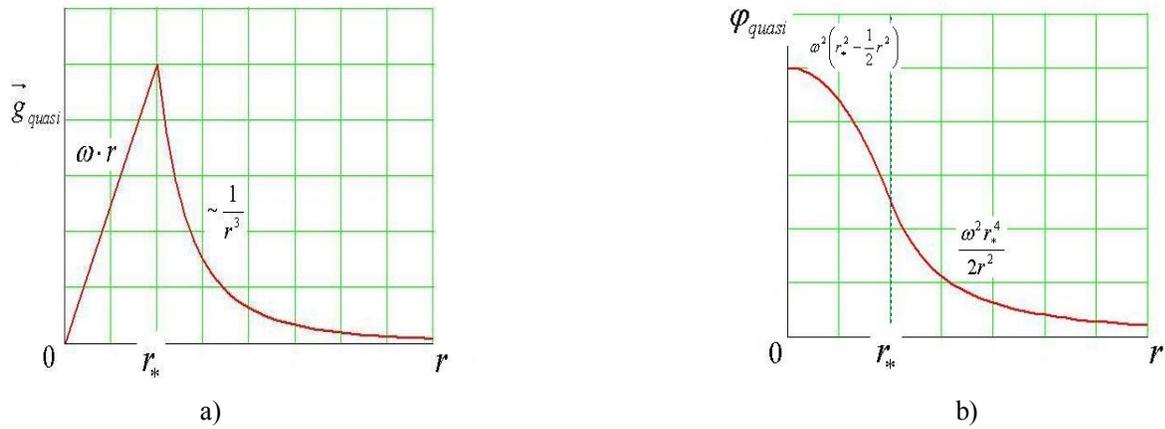

Fig.5.
Dependence of quasigravity field intensity (a) and potential (b) on a swirling flow radial coordinate $r$

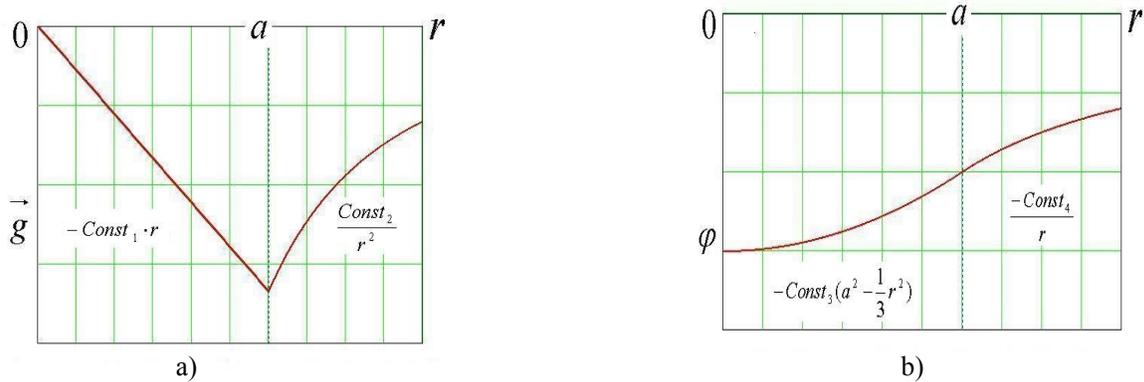

Fig.6.

Dependence of quasigravity field intensity (a) and potential (b) in a sphere with constant density substance $\rho = Const$

From the comparison of intensity and potential values we can conclude that both quasigravity and natural gravity fields are finite, but the former decrease more quickly $\left(g_{quasi} \sim 1/r^3\right)$. Inside a graviting sphere and in a gas flow core gravity fields demonstrate equal behaviour. Figure 5 and 6 show the comparison of intensities and potentials of natural and quasigravity fields for spherical and cylindrical symmetry. It should be noted that these fields have qualitative distinction: "genuine" gravity field potential is negative [1, 2], but quasigravity field potential has the positive sign.

The calculation shows that radial acceleration (quasigravity field intensity) in supersonic swirling flows of high-temperature plasma [7, 9] in special vortex devices can achieve giant values:

$$g_{quasi} = \frac{V_{\varphi max}^2}{r_*} \approx 10^8 \div 10^9 g. \tag{9}$$

Paradoxical is the fact that maximum velocity of gas rotation is many degrees less than the speed of light $V_{\varphi max} << c$, and the scalar curvature of a local space-time in a vortex with regard to (9) can amount to substantial values and has the negative meaning.

Considering the above said, it should be interesting to carry our a more detailed research of local space-time properties of the fields induced by strongly swirling gas flows on the basis of Einstein equations.

## III. Solution of Einstein equations for cylindrical symmetrical stationary swirling ideal gas flow with variable angular velocity and nonzero pressure

For axially symmetric stationary case of ideal gas rotation the metric can be written as follows:

$$dS^2 = g_{00}c^2dt^2 + g_{11}dr^2 + g_{22}d\varphi^2 + g_{33}dz^2 + 2g_{02}d\varphi dt + 2g_{13}d\varphi dr + 2g_{01}drdt. \tag{10}$$

In the interior zone of a swirling gas flow the law of a solid body rotation works $(\omega_\circ = Const)$, therefore the last two cross terms in (10) equal to zero. In the case of weak gravitation $g_{00} \approx g_{11} \approx g_{33} \approx 1$, $g_{22} \approx r^2$, and $g_{02} \approx \omega r^2$.

In the exterior zone of a swirling flow $(r > r_*)$ where angular acceleration is variable $\left(\omega \sim \frac{1}{r^2}\right)$, the last two cross terms in metric (10) remain. However, when sewing metrics in interior and exterior zones the corresponding elements of metric tensor $g_{ik}(\varphi_\Sigma)$ at the main coordinates $(r, \varphi, z)$ and their first derivatives must be equal

$$\begin{aligned} g_{\mu\nu}^{r_*-\varepsilon} &= g_{\mu\nu}^{r_*+\varepsilon}; \\ g'_{\mu\nu}{}^{r_*-\varepsilon} &= g'_{\mu\nu}{}^{r_*+\varepsilon}, \end{aligned} \quad \text{where } \varepsilon > 0 \tag{11}$$

Therefore, all three cross terms in the exterior zone also equal to zero.

Thus, the metric for cylindrical symmetrical field induced by swirling gas flows, in isotopic coordinates after nondimensionalization may be recorded in the following way:

$$d\bar{S}^2 = -\left(1+\frac{2\varphi}{c^2}\right)d\bar{t}^2 + \left(1-\frac{2\varphi}{c^2}\right)(d\bar{r}^2+d\bar{z}^2) + \left(1-\frac{2\varphi}{\tilde{n}^2}\right)\bar{r}^2 d\varphi^2 + \frac{2\omega_\circ r_*}{c}\left(1+\frac{2\varphi}{\tilde{n}^2}\right)\bar{r}^2 d\varphi d\bar{t}, \quad (10')$$

where $A = \frac{\omega_\circ r_*}{c}$, $\bar{r} = \frac{r}{r_*}$, $\bar{z} = \frac{z}{r_*}$, $\bar{t} = t\frac{c}{r_*}$, $\omega = \omega_\circ$ at $\bar{r} < 1$, $\omega = \frac{\omega_\circ}{\bar{r}^2}$ at $\bar{r} > 1$.

Due to nonlinear nature of Einstein equations the principle of fields superposition does not work. However, for weak gravity fields, for linearized Einstein equations the superposition principle is true. In the case of cylindrical or spherical symmetry in Newton approximation $g_{00}$ metric tensor element is equal to:

$$g_{00} = 1 + \frac{2\varphi_\Sigma}{c^2}, \quad (12)$$

where $\varphi_\Sigma$ - algebraic sum of potentials of a natural gravity field and a field induced by swirling gas flow. The total potential is equal to:

$$\varphi_\Sigma = \varphi_{body} + \varphi_{gas} + \varphi_{quasi}. \quad (13)$$

The sewing of natural gravity fields potentials is realized in a similar way (11) on the boundary of a massive body surface.

Einstein equations for simultaneous presence of a gravity field and a field induced by swirling gas flow can be put down in a mixed form for signature +2 in the following way:

$$R^\nu_\mu - \frac{1}{2}\delta^\nu_\mu R = -\frac{8\pi G}{c^4}T^\nu_\mu, \quad (14)$$

where $R^\nu_\mu$ - Ricci tensor, $\delta^\nu_\mu = 1$ ($\nu = \mu$), $\delta^\nu_\mu = 0$ ($\nu \neq \mu$), $R$ – Ricci tensor contraction, $G$ – gravity constant, $T^\nu_\mu$ - energy-momentum tensor.

It should be accentuated that quasigravity fields do not exist separately from natural gravity fields: quasigravity fields are local, because at a long distance they degenerate into natural gravity fields.

$$\vec{g}_{quasi} \Rightarrow \vec{g} = -\frac{MG}{r^2} + \frac{\omega^2 r_*^4}{r^3} \approx -\frac{MG}{r^2}$$

$$\varphi_{quasi} \Rightarrow \varphi = -\frac{MG}{r} + \frac{\omega^2 r_*^4}{2r^2} \approx -\frac{MG}{r} \qquad r > r_{**}$$

(8′)

where $M = M_{body} + M_{gas}$ : $M_{body}$ – is a mass of a certain body, in which gravity field a field of swirling gas flows is being studied; $M_{gas}$ - is a mass of gas in a swirling flow. At radius $r_{**}$ the total intensity of a swirling gas flow field and of a natural gravity field turns into zero (Fig.7).

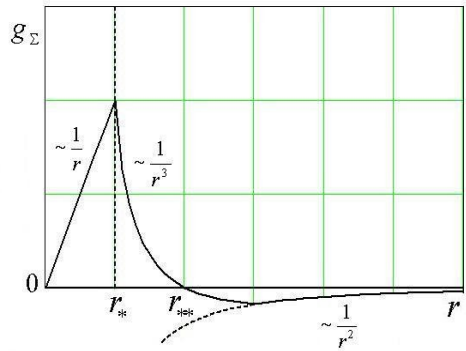

Fig.7.
Dependence of intensity $\vec{g}_\Sigma$ on radial coordinate with joint quasigravity and natural gravity fields action

$0 < r < r_{**}$ - area of quasigravity field domination

Below we will make calculations only for the case $|\rho_{quasi}| \gg \rho$, i.e. for the case of a local quasigravity field.

The calculation of elements of Riemann tensor and of energy-momentum tensor for the quasigravity field (14) with potentials (7-8) for exterior $r_* < r < r_{**}$ and interior $(r \leq r_*)$ cylinder area was carried out with the help of external differential forms technique [11] (for more details see Section IV). From Einstein equations on the basis of the calculated values of Ricci tensor elements $R_\mu^\nu$ (Section IV) we get at $\left(\frac{1}{c^2}\right)$ approximation the values of Ricci tensor $R_{\mu\nu}$, quasigravitation mass density $\rho_{quasi}$ :

$$\begin{cases} R_{\hat{t}\hat{t}} = 0 \\ R_{\hat{\chi}\hat{\chi}} = -3A^2 \\ R_{\hat{r}\hat{r}} = -3A^2 \\ R_{\hat{z}\hat{z}} = -2A^2 \end{cases} \qquad \rho_{quasi} = -\frac{\omega_\circ^2}{2\pi G}.$$

(15)

The contraction of Einstein equations for the studied field after simple transformation defines the field scalar curvature:

$$R = 8\pi G T_i^i / c^4 = -Q_G \cdot \gamma_1 \cdot \beta \cdot T^* \cong -8A^2, \qquad (16)$$

where $T^*$ - gas deceleration temperature; $\gamma_1 = 8\gamma$, $\gamma = \dfrac{2k}{k+1} \cdot \dfrac{1}{\mu}$ - parameter considering physic-chemical qualities of a swirling gas flow, inducing the field; $Q_G = R_G/c^2 \cong 0,92503 \cdot 10^{-13}$ $kg \cdot (Kmol \cdot K)^{-1}$ - constant of a field generated by swirling gas flows which presents the complex of fundamental universal constants; $k$ - Poisson constant; $R_G$ - universal gas constant; μ - molecular weight of a gas; $\lambda_{\varphi \max} = \lambda_1 r_1 / r_*$, $\beta = \lambda_{\varphi \max}^2$ - maximum peripheral velocity coefficient in the vortex; index "1" refers to parameters at the inlet.

As it follows from (15) the energy density of a quasigravity field is proportional to angular velocity squared which coincides with the conclusion of the research [13] that the higher is energy density of a vacuum field the higher is the speed of gas rotation.

From equations (16) it follows that scalar curvature of a studied field is in direct proportion to absolute gas temperature $T^*$ of the swirling flow.

In the exterior zone $r_* < r < r_{**}$ for metric (10) with potential (8) we get the following values of Ricci tensor elements (see IV).

$$\begin{cases} R_{\hat{t}\hat{t}} = \dfrac{4A^2}{\bar{r}^4} \\ R_{\hat{\chi}\hat{\chi}} = \dfrac{3A^2}{\bar{r}^4} \\ R_{\hat{r}\hat{r}} = -\dfrac{A^2}{\bar{r}^4} \\ R_{\hat{z}\hat{z}} = \dfrac{2A^2}{\bar{r}^4} \end{cases}, \qquad R = g^{\mu\nu} R_{\mu\nu} = 0 \qquad (17)$$

As the series expansion of metrical tensor expressions $g_{\mu\nu}$ in exterior zone $(r > r_*)$ is realized in powers $\left(\dfrac{1}{r}\right)$, than the expansion in powers of $\left(\dfrac{1}{r}\right)^4$ gives a good approximation to zero.

$$R_{\mu\nu} \approx \dfrac{Const}{r^4} \approx 0 \qquad (\mu = \nu). \qquad (18)$$

To calculate the value of quasigravity field pressure $p_{quasi}$ we use the equation of gravitational balance which is the consequence of the energy conservation law $\left(T^{\nu}_{\mu,\nu} = 0\right)$

$$\frac{\partial p_{quasi}}{\partial r} + \rho_{quasi}\varphi'_{quasi} = 0. \tag{19}$$

The expression for the local field pressure of gas flows $p_{quasi}$ is as follows:

$$p_{quasi} = \frac{\omega_\circ^4 r_*^2}{4\pi G}\left(1 - \bar{r}^{-2}\right). \tag{20}$$

The equation of quasigravity field state is:

$$\frac{p_{eq}}{\varepsilon_{eq}} = -\frac{\omega_\circ^2 r_*^2}{4c^2}\left(1 - \bar{r}^2\right) \tag{21}$$

or

$$\frac{p_{eq}}{\varepsilon_{eq}} = -\frac{\pi^2 \nu^2 r_*^2}{c^2}\left(1 - \bar{r}^2\right), \qquad \text{where} \quad \bar{r} = \frac{r}{r_*} \tag{21'}$$

Equations (21) and (21') may be called the optical form or "eikonal form" $\left(\omega_\circ^2/c^2\right)$ of local quasigravity field state equation.

After we express angular velocity of rotation $\omega$ through thermodynamic and kinematic parameters of a swirling gas flow, we can get another formula of state equation:

$$\frac{p_{quasi}}{\varepsilon_{quasi}} = -Q_G \cdot \gamma \cdot \theta^*\left(1 - \bar{r}^{-2}\right), \tag{22}$$

where $\theta^* = \lambda_{\theta\max}^2 \cdot T^*$ - is a value including thermodynamic $(p^*, T^*)$ and kinematic parameters of a gas.

Equation (21) is a gas-dynamic form of state equation for a quasigravity field.

The left part of the equation (21) includes quasigravity field parameters, yet the right part includes kinematic and thermodynamic parameters of a swirling gas flow that generates a quasigravity field.

Thus, the result obtained corresponds to the theorem (Ehlers, 1962) that to every statical vacuum solution there can be found a solid body rotational time independent solution for the dust.

The difference lies in the fact that in our case pressure is nonzero and the equation of state is more complicated than for the known vacuum solutions [14].

Besides, the qualitative difference of the equation of state (21, 22) from the known relativistic and ultra relativistic [14] is the fact that $0 < p_{eq} << \varepsilon_{eq}$ and it can be identified as infra relativistic case.

Thus, there has been obtained a new solution of Einstein equations for swirling gases with nonzero pressure and variable angular velocity. It refers to type I in Petrov's classification and according to the theorem (Ehlers, 1962) can be compared with statical vacuum solution of Einstein equations.

## IV. Particular Features of Quasigravity Fields Analysis

The calculation of Riemann $R^{\alpha}_{\beta\gamma\pi}$ and Ricci $R_{\mu\nu}$ tensors components of a quasigravity field will be performed via the method of exterior differential forms that were described in [11].

Metric (10) for $\bar{r} \leq 1$ after grouping of terms looks as follows:

$$d\bar{S}^2 = -(1+2A^2)d\bar{t}^2 + (1-2A^2+A^2\bar{r}^2)^2\bar{r}^2(d\varphi+\Omega d\bar{t})^2 + (1-2A^2+A^2\bar{r}^2)(d\bar{z}^2+d\bar{r}^2) \quad (10'')$$

where $\Omega = A\dfrac{(1+2A^2-A^2\bar{r}^2)}{(1-2A^2+A^2\bar{r}^2)} \cong A(1+4A^2-2A^2\bar{r}^2)$, $\varphi = \omega^2 r_*^2\left(1-\dfrac{\bar{r}^2}{2}\right)$, $d\chi = d\varphi+\Omega d\bar{t}$.

and describes space-time, generated by strongly swirling gas flows, i.e. quasigravity field.

Metric (10'') takes place in the area $0 < r < r_*$ where radial acceleration values $g_{quasi}$ are substantially bigger than natural gravity fields intensity.

Let us evaluate the doubled potential $2_\varphi$ and the metric tensor values $g_{\mu\nu}\left(g_{00}=a^2, g_{\mu\nu}=b, \mu=\nu=1,2,3\right)$ from potential (7, 8) and the given metric:

$$\begin{cases} a^2(\bar{r}) = 1+2A^2 \\ b^2(\bar{r}) = 1-\dfrac{2\omega^2\bar{r}_*^2 - \omega^2\bar{r}^2}{c^2} = 1-2A^2+A^2\bar{r}^2 \end{cases} \quad (23)$$

We can easily find orthonormal basis in this space-time:

$$dS^2 = -\left(\omega^{\hat{t}}\right)^2 + \left(\omega^{\hat{r}}\right)^2 + \left(\omega^{\hat{\chi}}\right)^2 + \left(\omega^{\hat{z}}\right)^2, \quad (24)$$

where

$$\begin{cases} \omega^{\hat{t}} = a(\bar{r})d\bar{t}; \\ \omega^{\hat{r}} = b(\bar{r})d\bar{r}; \\ \omega^{\hat{\chi}} = \bar{r}b(\bar{r})d\chi; \\ \omega^{\hat{z}} = b(\bar{r})d\bar{z}; \end{cases} \qquad \begin{aligned} a(\bar{r}) &= (1+2A^2)^{\frac{1}{2}}; \\ b(\bar{r}) &= \left(1 - \frac{A^2}{c^2} + \frac{k^2}{c^2}\bar{r}^2\right)^{\frac{1}{2}}. \end{aligned}$$

Calculation of connection

To find $\omega^{\hat{\alpha}}_{\hat{\beta}}$ we shall use the first Cartan equation:

$$d\omega^{\hat{\alpha}} + \omega^{\hat{\alpha}}_{\hat{\beta}} \wedge \omega^{\hat{\beta}} = 0;$$

$$d\omega^{\hat{t}} = a'd\bar{r} \wedge d\bar{t} = \frac{a'}{ab}\omega^{\hat{r}} \wedge \omega^{\hat{t}}; \qquad a' = \frac{\partial a}{\partial \bar{r}}; \qquad d\omega^{\hat{r}} = 0;$$

$$d\omega^{\hat{\chi}} = (b + \bar{r}b')d\bar{r} \wedge d\chi = \frac{\bar{r}b' + b}{\bar{r}b^2}\omega^{\hat{r}} \wedge \omega^{\hat{\chi}};$$

$$d\omega^{\hat{z}} = b'd\bar{r} \wedge dz = \frac{b'}{b^2}\omega^{\hat{r}} \wedge \omega^{\hat{z}}; \qquad \frac{b + \bar{r}b'}{\bar{r}b^2}\omega^{\hat{r}} \wedge \omega^{\hat{\chi}} + \omega^{\hat{\chi}}_{\hat{\alpha}} \wedge \omega^{\hat{\alpha}} = 0;$$

$$\frac{a'}{ab}\omega^{\hat{r}} \wedge \omega^{\hat{t}} + \omega^{\hat{t}}_{\hat{\alpha}} \wedge \omega^{\hat{\alpha}} = 0; \qquad \frac{b + \bar{r}b'}{\bar{r}b^2}\omega^{\hat{r}} \wedge \omega^{\hat{\chi}} = \omega^{\hat{r}} \wedge \omega^{\hat{\chi}}_{\hat{r}};$$

$$\frac{a'}{ab}\omega^{\hat{r}} \wedge \omega^{\hat{t}} = \omega^{\hat{r}} \wedge \omega^{\hat{t}}_{\hat{r}}; \qquad \omega^{\hat{\chi}}_{\hat{r}} = \frac{b + \bar{r}b'}{\bar{r}b^2}\omega^{\hat{\chi}} = \frac{b + \bar{r}b'}{b}d\chi.$$

$$\omega^{\hat{t}}_{\hat{r}} = \frac{a'}{ab}\omega^{\hat{t}} = \frac{a'}{b}d\bar{t};$$

$$\frac{b'}{b^2}\omega^{\hat{r}} \wedge \omega^{\hat{z}} + \omega^{\hat{z}}_{\hat{\alpha}} \wedge \omega^{\hat{\alpha}} = 0; \qquad \frac{b'}{b^2}\omega^{\hat{r}} \wedge \omega^{\hat{z}} = \omega^{\hat{r}} \wedge \omega^{\hat{z}}_{\hat{r}}; \qquad \omega^{\hat{z}}_{\hat{r}} = \frac{b'}{b^2}\omega^{\hat{z}} = \frac{b'}{b}d\bar{z}.$$

Thus, the complete set of connection forms $\omega^{\hat{\mu}}_{\hat{\nu}}$ looks like:

$$\begin{cases} \omega^{\hat{t}}_{\hat{r}} = \frac{a'}{ab}\omega^{\hat{t}} = \frac{a'}{b}d\bar{t}; \\ \omega^{\hat{\chi}}_{\hat{r}} = \frac{b + \bar{r}b'}{\bar{r}b^2}\omega^{\hat{\chi}} = \frac{b + \bar{r}b'}{b}d\chi; \\ \omega^{\hat{z}}_{\hat{r}} = \frac{b'}{b^2}\omega^{\hat{z}} = \frac{b'}{b}d\bar{z}. \end{cases} \qquad (25)$$

Calculation of quasigravity field curvature

Curvature is calculated by direct substitution of $\omega^{\hat{\mu}}_{\hat{\nu}}$ from correlation (25) in the curvature equation (the second Cartan equation)

$$\Omega^{\hat{\alpha}}_{\hat{\beta}} = d\omega^{\hat{\alpha}}_{\hat{\beta}} + \omega^{\hat{\alpha}}_{\hat{\gamma}} \wedge \omega^{\hat{\gamma}}_{\hat{\beta}}. \qquad (26)$$

From this with regard to (25) we get values of Cartan equation members (26) and values of Riemann tensors elements:

$$d\omega_{\hat{r}}^{\hat{t}} = \frac{a''b - a'b'}{b^2} d\bar{r} \wedge d\bar{t} = \frac{a''b - a'b'}{ab^3} \omega^{\hat{r}} \wedge \omega^{\hat{t}};$$

$$d\omega_{\hat{r}}^{\hat{\chi}} = \frac{(b' + b' + \bar{r}b'')b - b'(b + \bar{r}b')}{b^2} d\bar{r} \wedge d\chi = \frac{b'b + \bar{r}bb'' - \bar{r}(b')^2}{\bar{r}b^4} \omega^{\hat{r}} \wedge \omega^{\hat{\chi}};$$

$$d\omega_{\hat{r}}^{\hat{z}} = \frac{b''b - b'^2}{b^2} d\bar{r} \wedge d\bar{z} = \frac{b''b - b'^2}{b^4} \omega^{\hat{r}} \wedge \omega^{\hat{z}};$$

$$\Omega_{\hat{r}}^{\hat{t}} = d\omega_{\hat{r}}^{\hat{t}} + \omega_{\hat{\alpha}}^{\hat{t}} \wedge \omega_{\hat{r}}^{\hat{\alpha}} = \frac{a''b - a'b'}{ab^3} \omega^{\hat{r}} \wedge \omega^{\hat{t}}; \qquad (26')$$

$$R_{\hat{r}\hat{r}\hat{t}}^{\hat{t}} = \frac{a''b - a'b'}{ab^3} = -R_{\hat{r}\hat{t}\hat{r}}^{\hat{t}};$$

$$\Omega_{\hat{r}}^{\hat{\chi}} = d\omega_{\hat{r}}^{\hat{\chi}} + \omega_{\hat{\alpha}}^{\hat{\chi}} \wedge \omega_{\hat{r}}^{\hat{\alpha}} = \frac{b'b + \bar{r}bb'' - \bar{r}(b')^2}{\bar{r}b^4} \omega^{\hat{r}} \wedge \omega^{\hat{\chi}};$$

$$R_{\hat{r}\hat{r}\hat{\chi}}^{\hat{\chi}} = \frac{b'b + \bar{r}bb'' - \bar{r}b'^2}{\bar{r}b^4} = -R_{\hat{r}\hat{\chi}\hat{r}}^{\hat{\chi}};$$

$$R_{\hat{r}\hat{r}\hat{z}}^{\hat{z}} = \Omega_{\hat{r}}^{\hat{z}} = d\omega_{\hat{r}}^{\hat{z}} + \omega_{\hat{\alpha}}^{\hat{z}} \wedge \omega_{\hat{r}}^{\hat{\alpha}} = \frac{b''b - b'^2}{b^4} \omega^{\hat{r}} \wedge \omega^{\hat{z}};$$

$$R_{\hat{r}\hat{r}\hat{z}}^{\hat{z}} = \frac{b''b - b'^2}{b^4} = -R_{\hat{r}\hat{z}\hat{r}}^{\hat{z}};$$

$$\Omega_{\hat{\beta}}^{\hat{\alpha}} = d\omega_{\hat{\beta}}^{\hat{\alpha}} + \omega_{\hat{\gamma}}^{\hat{\alpha}} \wedge \omega_{\hat{\beta}}^{\hat{\gamma}};$$

$$R_{\hat{\chi}\hat{t}\hat{\chi}}^{\hat{t}} = \Omega_{\hat{\chi}}^{\hat{t}} = \omega_{\hat{\gamma}}^{\hat{t}} \wedge \omega_{\hat{\chi}}^{\hat{\gamma}} = \omega_{\hat{r}}^{\hat{t}} \wedge \omega_{\hat{\chi}}^{\hat{r}} = -\frac{a'}{ab} \cdot \frac{b + \bar{r}b'}{\bar{r}b^2} \omega^{\hat{t}} \wedge \omega^{\hat{\chi}};$$

$$R_{\hat{\chi}\hat{t}\hat{\chi}}^{\hat{t}} = -\frac{a'(b + \bar{r}b')}{\bar{r}ab^3};$$

$$R_{\hat{z}\hat{t}\hat{z}}^{\hat{t}} = \Omega_{\hat{z}}^{\hat{t}} = \omega_{\hat{\gamma}}^{\hat{t}} \wedge \omega_{\hat{z}}^{\hat{\gamma}} = \omega_{\hat{r}}^{\hat{t}} \wedge \omega_{\hat{z}}^{\hat{r}} = -\frac{a'}{ab} \cdot \frac{b'}{b^2} \omega^{\hat{t}} \wedge \omega^{\hat{z}}$$

$$R_{\hat{z}\hat{t}\hat{z}}^{\hat{t}} = -\frac{a'b'}{ab^3};$$

$$\Omega_{\hat{z}}^{\hat{\chi}} = d\omega_{\hat{z}}^{\hat{\chi}} + \omega_{\hat{\gamma}}^{\hat{\chi}} \wedge \omega_{\hat{z}}^{\hat{\gamma}} = \omega_{\hat{r}}^{\hat{\chi}} \wedge \omega_{\hat{z}}^{\hat{r}} = -\frac{b + \bar{r}b'}{\bar{r}b^2} \cdot \frac{b'}{b^2} \omega^{\hat{\chi}} \wedge \omega^{\hat{z}};$$

$$R_{\hat{z}\hat{\chi}\hat{z}}^{\hat{\chi}} = -\frac{b'(b + \bar{r}b')}{\bar{r}b^4};$$

$$R_{\hat{t}\hat{t}} = R_{\hat{t}\hat{r}\hat{t}}^{\hat{r}} + R_{\hat{t}\hat{\chi}\hat{t}}^{\hat{\chi}} + R_{\hat{t}\hat{z}\hat{t}}^{\hat{z}} = \frac{a''b - a'b'}{ab^3} + \frac{a'(b + \bar{r}b')}{\bar{r}ab^3} + \frac{a'b'}{ab^3} = \frac{\bar{r}a''b + a'b + \bar{r}a'b'}{\bar{r}ab^3};$$

$$R_{\hat{r}\hat{r}} = R^{\hat{t}}_{\hat{r}\hat{t}\hat{r}} + R^{\hat{\chi}}_{\hat{r}\hat{\chi}\hat{r}} + R^{\hat{z}}_{\hat{r}\hat{z}\hat{r}} = -\frac{a''b - a'b'}{ab^3} - \frac{b'b + \overline{r}bb'' - \overline{r}b'^2}{\overline{r}b^4} - \frac{b''b - b'^2}{b^4} =$$

$$= -\frac{\overline{r}b^2 a'' - \overline{r}ba'b' + ab'b + \overline{r}abb'' - \overline{r}ab'^2 + \overline{r}abb'' - \overline{r}ab'^2}{\overline{r}ab^4};$$

$$R_{\hat{\chi}\hat{\chi}} = R^{\hat{t}}_{\hat{\chi}\hat{t}\hat{\chi}} + R^{\hat{r}}_{\hat{\chi}\hat{r}\hat{\chi}} + R^{\hat{z}}_{\hat{\chi}\hat{z}\hat{\chi}} = -\frac{a'(b + \overline{r}b')}{\overline{r}ab^3} - \frac{b'b + \overline{r}bb'' - \overline{r}b'^2}{\overline{r}b^4} - \frac{b'(b + \overline{r}b')}{\overline{r}b^4} =$$

$$= -\frac{ba'(b + \overline{r}b') + ab'b + \overline{r}abb'' - \overline{r}ab'^2 + ab'b + a\overline{r}b'^2}{\overline{r}ab^4};$$

$$R_{\hat{z}\hat{z}} = R^{\hat{t}}_{\hat{z}\hat{t}\hat{z}} + R^{\hat{r}}_{\hat{z}\hat{r}\hat{z}} + R^{\hat{\chi}}_{\hat{z}\hat{\chi}\hat{z}} = -\frac{a'b'}{ab^3} - \frac{b''b - b'^2}{b^4} - \frac{b'(b + \overline{r}b')}{\overline{r}b^4} =$$

$$= -\frac{\overline{r}ba'b' + \overline{r}abb'' - \overline{r}ab'^2 + ab'(b + \overline{r}b')}{\overline{r}ab^4}.$$

Scalar curvature is evaluated as Ricci tensor trace:

$$R = R^t_t + R^r_r + R^\chi_\chi + R^z_z. \qquad (27)$$

Since we are investigating the case of weak gravity fields, in calculation of Ricci and Riemann tensor elements we may use the following calculation rules for the interior of domain $0 < r \leq r_*$:

$$\begin{cases} a \approx 1 + A^2; \quad b \approx 1 - A + \frac{A^2 \overline{r}^2}{2}; \\ a' = \frac{\partial a}{\partial r} = 0; \quad b' = \frac{\partial b}{\partial r} \approx \frac{A^2}{b}\overline{r}; \\ a'' = \frac{\partial^2 a}{\partial r^2} = 0; \quad b'' = \frac{\partial^2 b}{\partial r^2} = \frac{k^2\left(1 - \frac{A^2}{c^2}\right)}{c^2 b^3} \approx \frac{A^2}{b}. \end{cases} \qquad (28)$$

With regard to $\left(\frac{1}{c^2}\right)$ approximation we get values of Ricci, Riemann tensor elements and scalar curvature $R$:

$$\begin{cases} R^0_{110} = 0 \\ R^2_{112} = 2A^2 \\ R^3_{113} = A^2 \end{cases}, \quad \begin{cases} R^0_{202} = 0 \\ R^0_{303} = 0 \\ R^2_{323} = -A^2 \end{cases}, \quad \begin{cases} R_{00} = 0 \\ R_{11} = -3A^2 \\ R_{22} = -3A^2 \\ R_{33} = -2A^2 \end{cases};$$

$$R = g^{\mu\nu} R_{\mu\nu} \cong -8A^2, \tag{29}$$

where $t$, $r$, $\chi$, $z$ correspond to indices 0,1,2,3.

To determine the pressure and state equation in a quasigravity field we use the balance equation having calculated the pressure $p$ of a quasigravity field:

$$T^\nu_{\mu,\nu} = 0 \quad \Rightarrow \quad p_{,r} + \rho_{quasi} \varphi'_{,r} = 0 \quad \Rightarrow \quad p_{quasi} = -\frac{\omega^4 r_*^2}{4\pi G}\left(1 - \bar{r}^{-2}\right). \tag{19'}$$

The interval in the exterior domain $(\bar{r} > 1)$ is as follows:

$$d\bar{S}^2 = -\left(1 + \frac{2\varphi}{c^2}\right)d\bar{t}^2 + \left(1 - \frac{2\varphi}{c^2}\right)(d\bar{r}^2 + d\bar{z}^2) + \left(1 - \frac{2\varphi}{\tilde{n}^2}\right)\bar{r}^2 d\varphi^2 + \frac{2\omega(\bar{r})}{c}\left(1 + \frac{2\varphi}{c^2}\right)\bar{r}^2 d\varphi d\bar{t}$$

$$\varphi = \frac{\omega^2 \bar{r}_*^2}{2\bar{r}^2}, \quad \omega(\bar{r}) = \frac{\omega_\circ}{\bar{r}^2}.$$

After regrouping the equation members we get

$$d\bar{S}^2 = -\left(1 + \frac{2A^2}{\bar{r}^2}\right)d\bar{t}^2 + \left(1 - \frac{A^2}{\bar{r}^2}\right)(d\bar{r}^2 + d\bar{z}^2) + \left(1 - \frac{A^2}{\bar{r}^2}\right)\bar{r}^2\left[d\varphi + \frac{A}{\bar{r}^2}\frac{\left(1 + A^2/\bar{r}^2\right)}{\left(1 - A_2/\bar{r}^2\right)}dt\right]^2$$

The values of Reimann and Ricci tensors elements are calculated in the exterior domain by the following rules:

$$\begin{cases} a = (1 + 2A^2)^{\frac{1}{2}} \\ a \approx 1 + A^2 \\ b = (1 - 2A^2 + A^2\bar{r}^2)^{\frac{1}{2}} \\ a' = 0, \ a'' = 0 \\ b' = \frac{A^2\bar{r}}{b}, \ b'' = \frac{A^2}{b} - \frac{A^2}{b^2}A^2\bar{r}^2 \approx A^2 - A^4\bar{r}^2 \approx A^2 \end{cases} \qquad ab \approx 1; \quad \frac{1}{b} \approx a. \tag{31}$$

After simple transformation we determine the values of Reimann and Ricci tensors elements, using the technique of exterior differential forms (26').

$$\begin{cases} R^1_{010} = \dfrac{6A^2}{\bar{r}^4} = R_{1010} \\ R^2_{020} = -\dfrac{2A^2}{\bar{r}^4} = R_{2020} \\ R_{3030} = 0 \end{cases}, \qquad \begin{cases} R^2_{121} = R_{2121} = \dfrac{2A^2}{\bar{r}^4} \\ R^3_{131} = R_{3131} = \dfrac{3A^2}{\bar{r}^4} \\ R^3_{232} = R_{3232} = -\dfrac{A^2}{\bar{r}^4} \end{cases},$$

$$\begin{cases} R_{00} = \dfrac{4A^2}{\bar{r}^4} \\ R_{11} = -\dfrac{A^2}{\bar{r}^4} \end{cases}, \qquad \begin{cases} R_{22} = \dfrac{3A^2}{\bar{r}^4} \\ R_{33} = \dfrac{2A^2}{\bar{r}^4} \end{cases},$$

$$R \approx g^{\mu\nu} R_{\mu\nu} = 0, \tag{32}$$

as in the exterior domain the series expansion is realized in powers of $\dfrac{1}{r}$ and member $\dfrac{1}{r^4}$ may function as zero. The solution (32) shows that the space $(r > r_*)$ degenerate quickly into inane.

Weyl tensor for exterior $(\bar{r} > 1)$ and interior $(\bar{r} < 1)$ areas is defined in the following equation:

$$c_{iklm} = R_{iklm} - \dfrac{1}{2}\left(g_{il}R_{km} + g_{km}R_{il} - g_{il}R_{im} - g_{im}R_{kl}\right) + \dfrac{1}{6}\left(g_{il}g_{km} - g_{kl}g_{im}\right)R.$$

$$\begin{cases} c_{1010} = \dfrac{7A^2}{2\bar{r}^4} \\ c_{2020} = -\dfrac{5A^2}{2\bar{r}^4} \\ c_{3030} = -\dfrac{A^2}{\bar{r}^4} \end{cases}, \qquad \begin{cases} c_{2121} = \dfrac{A^2}{\bar{r}^4} \\ c_{3131} = \dfrac{5A^2}{2\bar{r}^4} \\ c_{3232} = -\dfrac{7A^2}{2\bar{r}^4} \end{cases}, \qquad \begin{cases} c_{00} = 0 \\ c_{11} = 0 \\ c_{22} = 0 \\ c_{33} = 0 \end{cases} c^i_i = 0$$

Nonzero values of Weyl tensor for $(\bar{r} < 1)$ equal to

The calculation check of nonzero Weyl tensor elements gives $g^{km} c_{iklm} = 0$ which proves the accuracy of calculation.

Let us analyse the features of space-time in exterior $(r > 1)$ and in terior $(r_* \leq 1)$ areas of a swirling gas. According to [2] to define the type of metric in Petrov's classification it is necessary to calculate invariants $I_1$, $I_2$ and Reimann tensor elements in contravariant and covariant forms.

Lowering and raising of indices in Reimann tensor is done by the following formulas:

$$R_{iklm} = g_{in} R^{n}{}_{klm};$$

$$\begin{cases} R^{1}{}_{010} = \dfrac{6A^2}{\bar{r}^4} \\ R^{2}{}_{020} = -\dfrac{2A^2}{\bar{r}^4}, \\ R^{3}{}_{030} = 0 \end{cases} \qquad \begin{cases} R^{2}{}_{121} = \dfrac{2A^2}{\bar{r}^4} \\ R^{3}{}_{131} = \dfrac{3A^2}{\bar{r}^4}, \\ R^{3}{}_{232} = -\dfrac{A^2}{\bar{r}^4} \end{cases}$$

$$\begin{cases} R_{1010} = \dfrac{6A^2}{\bar{r}^4} \\ R_{2020} = -\dfrac{2A^2}{\bar{r}^4}, \\ R_{3030} = 0 \end{cases} \qquad \begin{cases} R_{2121} = \dfrac{2A^2}{\bar{r}^4} \\ R_{3131} = \dfrac{3A^2}{\bar{r}^4} \\ R_{3232} = -\dfrac{A^2}{\bar{r}^4} \end{cases}$$

Raising of indices of Ricci tensor is done by the formula

$$R^{ik}{}_{lm} = g^{kn} R^{i}{}_{nlm};$$

$$\begin{cases} R^{1010} = \dfrac{6A^2}{\bar{r}^4} \\ R^{2020} = -\dfrac{2A^2}{\bar{r}^4}, \\ R^{3030} = 0 \end{cases} \qquad \begin{cases} R^{2121} = \dfrac{2A^2}{\bar{r}^4} \\ R^{3131} = \dfrac{3A^2}{\bar{r}^4}, \\ R^{3232} = -\dfrac{A^2}{\bar{r}^4} \end{cases}$$

$$\begin{cases} R^{10}{}_{10} \cong -\dfrac{6A^2}{\bar{r}^4} \\ R^{20}{}_{20} \cong \dfrac{2A^2}{\bar{r}^4}, \\ R^{30}{}_{30} = 0 \end{cases} \qquad \begin{cases} R^{21}{}_{21} = \dfrac{2A^2}{\bar{r}^4} \\ R^{31}{}_{31} = \dfrac{3A^2}{\bar{r}^4} \\ R^{32}{}_{32} = -\dfrac{A^2}{\bar{r}^4} \end{cases}$$

$$I_1 = \frac{1}{48}\left[ R_{1010} R^{1010} + R_{2020} R^{2020} + R_{3030} R^{3030} + R_{2121} R^{2121} + R_{3131} R^{3131} + R_{3232} R^{3232} \right] =$$

$$= \frac{1}{48}\left[ \frac{6A^2}{\bar{r}^4} \cdot \frac{6A^2}{\bar{r}^4} + \frac{4A^4}{\bar{r}^8} + 0 + \frac{2A^2}{\bar{r}^4} \cdot \frac{2A^2}{\bar{r}^4} + \frac{3A^2}{\bar{r}^4} \cdot \frac{3A^2}{\bar{r}^4} + \frac{A^4}{\bar{r}^8} \right] = \frac{A^4}{48\bar{r}^8}[36 + 4 + 4 + 9 + 1] = \frac{54}{48}\frac{A^4}{\bar{r}^8} = \frac{9}{8}\frac{A^4}{\bar{r}^8}$$

.

$$I_2 = \frac{1}{96}\left[ R^{iklm}_{1010} R^{1010} R^{10}_{..10} + R_{2020} R^{2020} R^{20}_{..20} + R_{3030} R^{3030} R^{30}_{..30} + R_{3131} R^{3131} R^{31}_{..31} + R_{2121} R^{2121} R^{21}_{..21} + R_{3232} R^{3232} R^{32}_{..32} \right] =$$

$$= \frac{1}{96}\left[ \frac{36A^4}{\bar{r}^8}\left(-\frac{6A^2}{\bar{r}^4}\right) + \frac{4A^4}{\bar{r}^8}\frac{2A^2}{\bar{r}^4} + 0 + \frac{9A^4}{\bar{r}^8}\frac{3A^2}{\bar{r}^4} + \frac{A^4}{\bar{r}^8}\left(-\frac{A^2}{\bar{r}^4}\right) \right] = \frac{A^6}{\bar{r}^{12} 96}[-218 + 8 + 27 - 1] = -\frac{23}{12}\frac{A^6}{\bar{r}^{12}}$$

.

$$I_1 = \frac{1}{48}\left( R_{iklm} R^{iklm} - i R_{iklm} \overset{*}{R}{}^{iklm} \right),$$

$$I_2 = \frac{1}{96}\left( R_{iklm} R^{lmpr} R^{ik}_{pr} + i R_{iklm} R^{impr} \overset{*}{R}{}^{ik}_{pr} \right),$$

where $i$ - unit imaginary number, $\overset{*}{R}_{iklm} = \frac{1}{r} E_{ikpr} R^{pr}_{lm}$ - dual tensor for $R_{iklm}$, $E^{ikpr} = \frac{1}{\sqrt{-g}} e^{iklm}$;

$g$ - metric tensor $g_{\mu\nu}$ determinant, $e^{iklm}$ - absolutely non-symmetrical pseudotensor.

In our case to determine the invariants $I_1$, $I_2$ it is enough to calculate their real parts as dual tensors $\overset{*}{R}{}^{iklm}$ and $\overset{*}{R}{}^{ik}_{pr}$ identically go to zero. Absolutely non-symmetrical tensor, $e^{ikik} \equiv 0$, then the invariants $I_1$ and $I_2$ look as follows:

$$I_1 = \frac{1}{48}\left( R_{iklm} R^{iklm} \right),$$

$$I_2 = \frac{1}{96}\left( R_{iklm} R^{lmpr} R^{ik}_{pr} \right).$$

Then tensor invariant

$$I_1^3 - 27 I_2^2 = M$$

will be nonzero and real which allows to define metric classification in exterior gas space $(r > r^*)$ as type I according to Petrov.

Due to the determination of $I_1$ and $I_2$ in [18] the criteria of metric type classification is

$$M = I_1^3 - 6 I_2^2.$$

In our case $M = \frac{A^{12}}{\bar{r}^{24}}\left(\frac{729}{512} - 27\frac{529}{144}\right) < 0$, i.e. $M$ is negative, which allows to classify the metric as type $I(M^-)$.

### V. Conclusions and Prospects of Further Research

1. Using theoretical and experimental results the present research shows that there is a class of noninertial frames of reference which satisfy equivalency principle more than the known noninertial systems – these are strongly swirling gaseous flows.

It is shown, that the intensity and potential of a field generated by strongly swirling gaseous flows demonstrate the behaviour qualitatively similar to that of the natural gravity fields, they degenerate in infinity to Galilean space, but decrease faster than the natural gravity fields, correspondingly $\varphi_{quasi} \sim \dfrac{1}{r^2}$, $\vec{g} \sim \dfrac{1}{r^3}$ and $\varphi \sim \dfrac{1}{r}$, $\vec{g} \sim \dfrac{1}{r^2}$.

The potential and pressure of a field generated by strongly swirling gas flows have the positive sign, i.e. opposite to that of a natural gravity field.

2. As the analysis showed, under certain operating and geometrical conditions in special swirl chambers the centrifugal acceleration and therefore, intensity $\vec{g}_{quasi}$ can mount to giant values $g_{quasi} \approx 10^9 - 10^{10}\, м/с^2$. Paradoxical is the fact that the maximum velocity of gas rotation $(V_{\varphi \max} \sim 10^4\, м/с)$ is many degrees less than the speed of light $c\,(V_\varphi \ll c)$, and the corresponding scalar curvature of a local space generated by a strongly swirling gas flow can amount to substantial values ($R_{quasi} \sim -10^{-4} \div 10^{-5}\, 1/м^{-2}$).

It was shown that a scalar curvature of a quasigravity field is directly proportional to the absolute temperature of a gas: $R_{quasi} \sim T^*$.

3. There has been obtained a solution of Einstein equations for stationary cylindrical symmetrical swirling ideal gas flows with nonzero pressure and variable angular velocity which refers to type I in Petrov's classification.

4. State equation of quasigravity field (23, 24) imposes the negative connection between pressure $p_{quasi}$ and energy density $\varepsilon_{quasi}$, as well as in the case of vacuum solutions of Einstein equations, but the connection between the noted parameters of a quasigravity field is more complicated (it depends on physic-chemical and thermodynamic parameters of a swirling gas flow).

5. The convolution of Einstein equations for a quasigravity field made it possible to single out a new constant: the complex of the universal gas constant and light $\left( \theta_G = \dfrac{R}{c^2} \approx 0{,}9250539 \cdot 10^{-13}\, \dfrac{kg}{Kmol \cdot K} \right)$.

The analysis showed that quasigravity fields cannot exist without natural gravity fields, they are local and at sufficient distance $(r \gg r_*)$ they degenerate in natural gravity fields.

Quasigravity fields can be artificial as well as natural. As a rule, artificial quasigravity fields are small-scale formations – these are fields created by strongly swirling gas or plasma flows in vortex and whirlwind chambers. The scale of natural quasigravity fields may vary over wide range. Natural whirlwinds are medium-scale quasigravity formations $(1 \div 10^3$ м$)$, and flows of a swirling gaseous substance in the accretational disks of black holes, flows of gas about galactic nucleus [13] are of astronomical scale.

The prospects of further research of quasigravity fields are first of all connected with the experimental confirmation of the suggested theory.

As the calculations show, while passing a quasigravity field generated by a swirling flow of high-temperature plasma, relative changing of photon radiation frequency can amount the following values:

$$\frac{\Delta \nu}{\nu} = \frac{\int_0^{r^*+\varepsilon} g_{quasi} dr}{c^2} \approx 10^{-10} \div 10^{-11}. \tag{33}$$

At that photon should "become blue", because the potential of a quasigravity field is positive, unlike that of a natural gravity field.

As early as in 1960 R.V. Pound and G.A. Rebka [16] conducted a direct experiment to verify the theory of relativity: they measured relative change of photon radiation frequency in the field of the Earth gravitation (h=21m).

$$\frac{\Delta \nu}{\nu} = \frac{gh}{c^2} \approx 10^{-15} \tag{34}$$

to 1% accuracy.

Contemporary optical and quantum methods allow verifying the predicted values of relative photon frequency changes while they pass through a quasigravity field (33).

Promising directions of the further research are theoretical and experimental investigations of the strong quasigravity fields that are realized, for example, in synchrotrons, where velocities of electrons rotation can achieve $V_\theta \approx 0{,}95c$ [17]. In this case quasigravity field potential $\varphi_{quasi}$ and intensity $g_{quasi}$ can amount to the giant values and may be determined by the following expressions:

$$\varphi_{quasi} = \frac{B^2 R^2}{2}\left(\frac{e}{m_e}\right)^2 \approx \frac{10^{17}}{2} \approx \frac{1}{2}10^{17}\frac{м^2}{c^2};$$

(35)

$$g_{quasi} = B^2 R\left(\frac{e}{m_e}\right)^2 \approx \frac{10^{17}}{0,1} \approx 10^{18}\frac{м}{c^2},$$

where $e$, $m_e$ – charge of the electron and its mass correspondingly, $B$ – magnetic induction, $R$ – radius of a synchrotron channel.

At this relative changing of photon frequency can come up to values $\frac{\Delta \nu}{\nu} \sim 1$.

**Acknowledgements**

In conclusion the author would like to thank Professor V.G. Shakhov for his constructive remarks on the article, Associate professor D.F. Kitayev, A.M. Sukhov for assistance in using the exterior differential forms apparatus, and S.L. Safronova, T.A.Bazhenova and E.I.Boitsova for technical assistance.